\documentclass[a4paper,11pt]{article}
\usepackage{jinstpub} 
\usepackage{etoolbox}
\usepackage{setspace}
\usepackage{subcaption}
\usepackage{float}
\usepackage{amsmath}
\usepackage{svrsymbols}
\usepackage{titletoc}
\usepackage[titles]{tocloft}
\usepackage{tikz-feynman}
\usepackage{xcolor}
\usepackage{graphicx}
\tikzfeynmanset{compat=1.1.0}
\usepackage{hyphenat}
\usepackage{longtable}
\usepackage{natbib} 
\usepackage{setspace} 

\title{\boldmath New lower background and higher rate technique for anti-neutrino detection using Tungsten 183 Isotope}

\author{J. Novak, N. Solomey$^1$\note{Corresponding author}, B. Hartsock, B. Doty, and J. Folkerts}
\affiliation{Wichita State University,\\Physics Division of the Dept. of Mathematics, Statistics and Physics\\
1845 Fairmount St, Wichita, KS, USA}
\emailAdd{nick.solomey@wichita.edu}

\abstract{Low energy anti-neutrinos detected from reactors or other sources have typically used the conversion of an anti-neutrino on Hydrogen, producing a positron and a free neutron. This neutron is subsequently captured on a secondary element with a large neutron capture cross-section such as gadolinium or cadmium. 
With most neutron captures on gadolinium, it is possible to get two or three delayed gamma signals of known energy to occur. Modern experiments can make measurements with timing on the order of 25 ns. 
Fast electronics like these allow for the possibility of accessing the very fast signals from the nuclear de-excitation of a heavy nucleus following the prompt positron signal, rather than relying on traditional IBD techniques.
We have found an isotope of tungsten, $^{183}$W that produces tantalum in the ground state at 2.094 MeV or the first excited state at 2.167 MeV. The excited state of $^{183}$Ta$^*$ emits a signature secondary gamma pulse of 73 keV with a 106 ns half-life. This offers a new delayed coincidence technique that can be used to identify anti-neutrinos with lower background noise. This allows for less shielding than required for modern inverse beta decay detectors.}


\arxivnumber{2311.08418} 

\begin{document}
\maketitle
\flushbottom

\section{Introduction}\label{sec:Intro}
\begin{figure}[htbp]
\centering
\includegraphics[width=.6\textwidth]{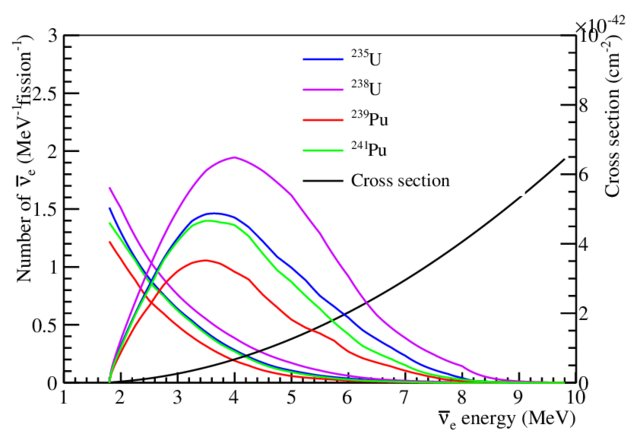}
\caption{Number of neutrinos per MeV of fission in a typical nuclear reactor, cross section for inverse beta decay, and their convolution \cite{Juno}. Each nucleus includes the neutrinos from all daughters in its chain.
\label{reactorneutrinos}}
\end{figure}

Historically the first direct detection of the neutrino was using anti-neutrinos from a nuclear reactor by Cowan and Reines in the 1950s \cite{Cowan}. Nuclear reactors produce many different energies of anti-neutrinos, not just from the fission process itself, but also from the daughter nuclei produced in the fission, see Figure \ref{reactorneutrinos} \cite{Juno}. 
The initial focus of this research was to determine if the monitoring of the anti-neutrino rate from nuclear reactors, specifically within submarines, from the anti-neutrinos emitted was possible, and to create a better method of detection for defense applications.
Our group has recently been working under NASA grants to design a solar neutrino detector. That work uses a double pulse signal from a gallium interaction, which was seminal in the studies of this paper. To improve detection of reactor anti-neutrinos, rather than only using the double coincidence approach used in the detection of solar neutrinos, we have studied elements which could produce 2 or more coincidence signals with anti-neutrinos, which led us to tungsten 183. 

\begin{figure}[htbp]
\centering
\includegraphics[width=.4\textwidth]{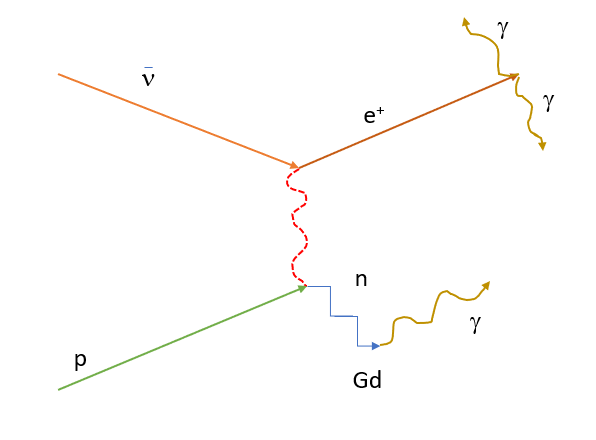}
\caption{Anti-neutrino interaction with a proton, producing a positron annihilation signal and a neutron. Once the neutron has thermalized, it can be absorbed and one or more photons are emitted in the decay of the resultant excited nucleus.
\label{fix1}}
\end{figure}

The standard process of anti-neutrino detection\cite{Cowan}, depicted in Figure \ref{fix1}, is a weak-nuclear force process of an anti-neutrino interacting with a proton:
\begin{equation*}
    \bar{\nu_{e}} + p^+ \rightarrow e^+ + \hspace{1mm} ^0n
\end{equation*}
where the positron annihilates on an atomic shell electron. This makes a distinct signal from the positron annihilation, but the neutron is fast and must slow down to be captured. The neutron's passage through matter slows it to capturable energies, but it takes $\sim$ 200 $\mu$s \cite{nTiming}. When the neutron is absorbed on a heavy element such as gadolinium or cadmium, the resultant excited nucleus emits many different possible cases of secondary
gammas when it decays, see Table \ref{neutrongammas} for gadolinium. In the case of gadolinium one such process only involves the emission of a single gamma and the other 11 have two or more gammas emitted with delayed timings between them. This double coincidence was measured in the 1950s with the slow old electronics of that era ($\sim 10$ $\mu$s \cite{Cowan}), but the large >200 $\mu$s signal delay made that measurement possible. Today with fast electronics it is possible to consider using a double or triple delayed coincidence with smart electronics selection to reject the out of time background signals. These electronics can be combined with fast scintillators. In contrast with the interaction on a lone proton, if the anti-neutrino has enough energy this reaction can occur on a proton in a heavy element.

A key idea studied here 
was a comprehensive survey of the known atomic isotopes to find the most promising candidates. 
We calculated anti-neutrino energy threshold and secondary emission parameters for a particularly promising process, which is described in section \ref{sec:Tungsten}. 
We conclude in section \ref{sec:Conclusion} with a discussion of possible applications for this technology, including scientific, economic, and national security.

\begin{table}[htbp]
\centering
\caption{
Known primary, secondary, and tertiary gamma rays produced from neutron capture on gadolinium with their relative intensities. Energies are given in keV \cite{Tanaka}.
\label{neutrongammas}}
\smallskip
\begin{tabular}{|rcll|}
\hline
Primary & Secondary & Tertiary & Intensity [\% 10$^{-2}$] \\
\hline
5661 & 2786 & - & 15.4\\
5698 & 2749 & - & 28.6\\
5779 & 2672 & - & 18.8\\
5885 & 2563 & - & 9.0\\
5885   & 2364 & 199 & 8.4\\
6034 & 2412 & - & 14.0\\
6034   & 2213 & 199 & 6.4 \\
6319 & 2127 & - & 9.4 \\
6348 & 2188 & - & 12.1 \\
6348 & 2097 & 199 & 9.8 \\
6348 & 1036 & 1154 & 4.6 \\
6348 & 1036 & 1065 & 3.8 \\
6430 & 2017 & - & 20.7 \\
6430 & 1818 & 199 & 11.7 \\
6474 & 1964 & - & 35.2 \\
7288 & 1158 & - & 34.8 \\
7288 & 959 & 199 & 10.5 \\
7382 & 1154 & - & 12.7 \\
7382 & 1065 & - & 10.6 \\
8448 & - & - & 1.8 \\
\hline
\end{tabular}
\end{table}

\newpage

\section{Use of a tungsten isotope as a unique anti-neutrino detector medium}\label{sec:Tungsten}
When compared with neutrino detectors, the accidentals in anti-neutrino detectors are generally small due to the characteristic prompt positron signal followed by the $\sim200\text{ }\mu$s delayed neutron signal. This double-pulse signal allows many anti-neutrino inverse beta decay (IBD) detectors to be built with less passive shielding than their neutrino counterparts. In this section, we will discuss why tungsten is such an interesting candidate, and how the use of tungsten could allow for a much tighter timing window, which would improve the background rejection and increase sensitivity in turn. 

\begin{table}[htp!]
\begin{center}
\caption{Table of the only nuclear excited state capable of producing a double timing pulse from the reaction of low energy reactor anti-neutrinos with tungsten, $^{183}$W, energy threshold for the process, their decay product gamma energy, and half-life \cite{ref:NuclearDataSheets}}
\begin{tabular}{ | c | c | c | c | }
\hline
{Reaction Products} & {Energy Threshold} & {Photon Energy} & {Half Life} \\ 
 \hline
Ta$_{73}^{183M1}+e^+$& 2.167 MeV & 0.073 MeV & 106 ns \\
 \hline
\end{tabular}
\label{tab:AntiNeutrinoReactions}
\end{center}
\end{table}

\begin{figure}[htbp]
\centering
\includegraphics[width=.4\textwidth]{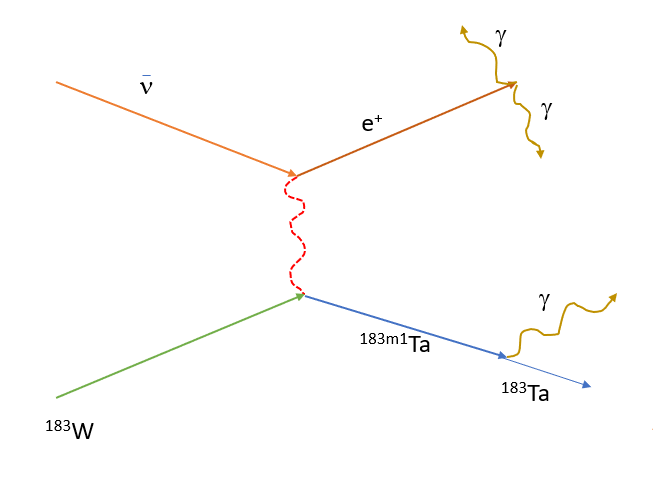}
\caption{Anti-neutrino interaction off of a proton, see Table \ref{tab:AntiNeutrinoReactions}. 
\label{fix2}}
\end{figure}

Anti-neutrino detection would require a scintillator with a large amount of tungsten, and $^{183}$W is only 14\% of naturally occurring tungsten. Looking for possible scintillators, we find that not many use tungsten. The few which do tend to be low light yield, like PbWO\textsubscript{4}, are used to detect Cerenkov light and are not sensitive to the neutron signal required for the double pulse signature of IBD. This is changing due to recent developments with glass scintillators. In particular tungsten gallium-phosphate glasses, such as NaPGaW, have emerged with the necessary material properties needed to perform a double pulse detection\cite{Lodi}. These glasses do not contain a neutron capture target, but they could be used to look for the tantalum de-excitation gamma. These glasses also have the advantage that they can be doped with a 10\% mol concentrate of $^{183}$W. That work created samples in several concentrations, but it is unclear if 10\% was the limit or where they chose to stop\cite{Lodi}.

By using the NaPGaW glass scintillator, we are able to control the specific tungsten isotope, which in turn gives us the ability to use this new detection method of $^{183}$W to $^{183 m1}$Ta$^*$ as shown in Figure \ref{fix2}. This means we can use the excited state gamma as the second pulse after the positron annihilation, rather than the neutron capture. With the 106 $ns$ half-life, the double pulse coincidence happens quickly enough that the window for background noise is decreased significantly. Such reductions in background noise can be found in gallium neutrino detectors using an interaction with a very similar half-life \cite{ref:NoiseReduction}. 
The tungsten excited-state reaction does have an upper energy range of $\approx$ 4.5 MeV, due to how weakly bounded $^{183}$W is. If $^{183}$W absorbs more than 4.5 MeV of energy, that energy can induce fission in the tungsten nucleus. This process is a type of induced fission that occurs predominantly in extremely heavy atomic nuclei. In this event, a single nucleus divides into two or more smaller nuclei, accompanied by the emission of several neutrons and a significant release of energy. Whether or not the isotope is going to be within the induced fission range depends on how much energy the tungsten nucleus takes after the interaction. 
This induced fission will create a signal very different from the nuclear de-excitation. Because of the energy and heavy nuclei involved, it should be possible to use this signal to also detect anti-neutrinos in the same detector with different signal cuts.
Further studies need to be performed to determine the cross-section and nuclear form factors of tungsten as well as the amount of energy in the induced fission reaction \cite{Novak_MS}.

Any of the proposed detectors using a tungsten-based scintillator could be improved by adding a thin film of Gd, perhaps using it as the reflector on the scintillator. This would allow a tungsten-based detector to still access the IBD signal. In total, there are now three methods of detection with $^{183}$W. Such a detector could be designed to look for the lower-rate neutron capture signal, the higher rate $^{183}$W double pulse signal, and the instantaneous induced fission signal. A hybrid detector with smaller segmented Gadolinium Aluminum Gallium Garnet (GAGG) crystals that have $^{183}$W glued between the crystals in thin layers is also an acceptable solution which allows access to the neutron capture signal.

\begin{figure}[htbp]
    \centering
    \begin{subfigure}{0.45\textwidth}
        \centering
        \includegraphics[width=\textwidth]{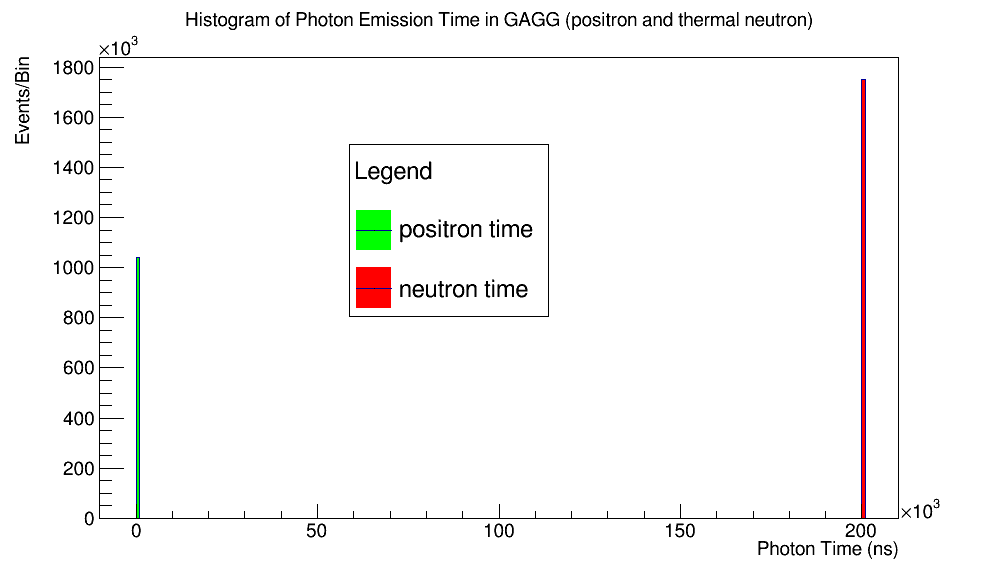}
        \caption{}
        \label{fig:IBD-double}
    \end{subfigure}
    \hfill 
    \begin{subfigure}{0.45\textwidth}
        \centering
        \includegraphics[width=\textwidth]{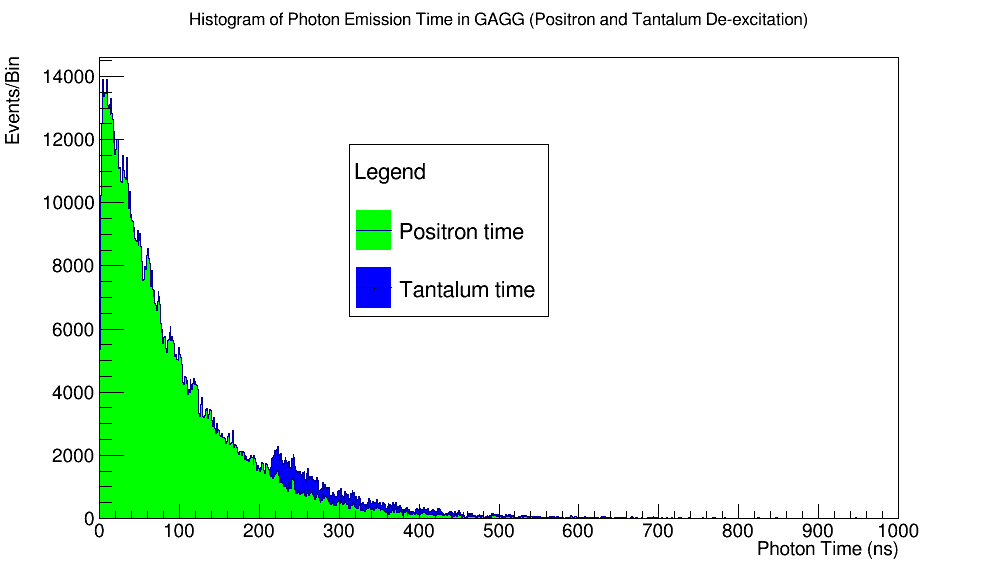}
        \caption{}
        \label{fig:Tantalum-double}
    \end{subfigure}
    \caption{(a) Photon emission timing histogram of a
 single simulated inverse beta decay event with a prompt positron (left peak) and a $\sim200$ $\mu$s delayed thermal neutron (right peak). The simulated detector is a GAGG crystal surrounded by tungsten. Exponential GAGG signals appear as in a single bin due to the fast decay timing of the scintillator. (b) Photon emission timing from a simulated prompt positron and delayed tantalum de-excitation gamma from the tungsten and anti-neutrino interaction. An interaction with the de-excitation gamma at 2 half lives from the initial positron was chosen for human readability.}
    \label{fig:double-pulse}
\end{figure}

Simulations of a detector made from GAGG surrounded by tungsten shows that a fast scintillator like GAGG is capable of resolving inverse beta decay double pulse signals, see Figure \ref{fig:double-pulse}\cite{Hartsock}. These simulations, in combination with the calculations above, give us confidence that a tungsten-based detector is a promising new technology. To push this technology, we are planning to conduct laboratory studies and experiments to constrain the uncertain aspects of a $^{183}$W detector.

\section{Applications and Conclusion}\label{sec:Conclusion}
As stated in Section \ref{sec:Intro}, we started this research wanting to improve anti-neutrino detection methods for purposes such as national defense and locating nuclear submarines. The results indicate that $^{183}$W is a good candidate for a lower noise detector of anti-neutrinos, and a detector with a sufficiently high cross-section could possibly detect nuclear reactors at large distances \cite{Novak_MS}. While this was the starting point, further research of this isotope has resulted in a wide variety of applications.

An anti-neutrino detector using an isotope that produces a short-lived excited state could be highly-sensitive and low-background.
Our technique using $^{183}$W for anti-neutrinos can be used to improve science with reactor events \cite{Novak_MS}. This anti-neutrino detection technique could also be used to improve Geo-neutrino detection. This has applications in natural resource searches and probing geological structure, which is of interest both on earth and on nearby bodies like the moon, Venus, asteroids, and other earth-like planets.

\acknowledgments
This work from 2021-2023 is the Master of Science in Physics Thesis by J. Novak at Wichita State University under the guidance and the introduction of the new idea for excited state tungsten by Nick Solomey serving as the Thesis adviser.

\bibliographystyle{JHEP}
\bibliography{biblio}

\end{document}